\documentclass[aps,prl,reprint,superscriptaddress,floatfix,twocolumn,nofootinbib]{revtex4-1}
\usepackage{graphicx}
\usepackage{amsmath}
\usepackage{amsfonts}
\usepackage{bm}
\usepackage{color}
\usepackage{indentfirst}
\usepackage[normalem]{ulem}
\begin{document}
\title{Dynamically Heterogeneous Relaxation of Entangled Polymer Chains}

\author{Yuecheng Zhou}

\affiliation{Department of Materials Science and Engineering, University of Illinois at Urbana-Champaign, Urbana, Illinois 61801, United States}

\author{Charles M. Schroeder}
\affiliation{Department of Materials Science and Engineering, University of Illinois at Urbana-Champaign, Urbana, Illinois 61801, United States}
\affiliation{Department of Chemical and Biomolecular Engineering, University of Illinois at Urbana-Champaign, Urbana, Illinois 61801, United States}

\email{cms@illinois.edu}

\date{\today}
\begin{abstract}
	Stress relaxation following deformation of an entangled polymeric liquid is thought to be affected by transient reforming of chain entanglements. In this work, we use single molecule techniques to study the relaxation of individual polymers in the transition regime from unentangled to entangled solutions. Our results reveal the emergence of dynamic heterogeneity underlying polymer relaxation behavior, including distinct molecular sub-populations described by a single-mode and a double-mode exponential relaxation process. The slower double-mode timescale $\tau_{d,2}$ is consistent with a characteristic reptation time, whereas the single-mode timescale $\tau_s$ and the fast double-mode timescale $\tau_{d,1}$ are attributed to local regions of transient disentanglement due to deformation. 
\end{abstract}

\maketitle
	Entangled polymeric liquids are ubiquitous in materials processing and have garnered broad interest in condensed matter physics for many years \cite{DeGennes1976a}. Topological constraints in entangled polymer solutions and melts result in a dramatic slow down in chain dynamics, which is commonly modeled using the classic tube theory by de Gennes \cite{DeGennes1976a} and Doi and Edwards \cite{Doi1986}. The tube model relies on a mean-field approximation by considering a single polymer chain moving or reptating through a confinement potential due to obstacles created by neighboring chains \cite{DeGennes1976b, Rubinstein2003}. Recent work has considered topological entanglements in a self-consistent manner at the level of microscopic forces \cite{Schweizer2016}, which avoids the ad hoc assumptions of a confining tube while fundamentally deriving an effective confinement potential for entanglements.
    
    A fundamental question underlying polymer solutions and melts focuses on how stress relaxes in topologically entangled systems. Following a large deformation, the original Doi-Edwards model (D-E) assumes that polymers undergo a fast chain retraction along the confining tube, followed by a slow stress relaxation via reptation to relax non-equilibrium orientations due to the deformation. Although the original D-E model was successful in capturing some aspects of the physics of entangled polymer solutions \cite{Lodge1990,McLeish2002a}, experimentally determined longest relaxation times $\tau_d$ for melts exhibit a molecular weight $M$ dependence of $\tau_d \sim M^{3.4}$  \cite{Ferry1980, Lodge1999, Larson1999}, whereas the D-E model predicts a molecular weight dependence of $M^3$. To reconcile this discrepancy, the original tube theory was extended to include constraint release (CR) \cite{Graessley1982,Viovy1991} and contour length fluctuations (CLF) \cite{Doi1983,Milner1998}. For large non-linear deformations, the D-E model was further extended to account for chain stretching (CS) \cite{Marrucci1988} and convective constraint release (CCR) to account for dynamic release of entanglements in flow. Accurate treatment of these phenomena resulted in an advanced microscopic theory (GLaMM) that captures a wide range of dynamic properties of entangled polymers \cite{Graham2003}.
	
	In addition to theoretical modeling, entangled polymer solutions have been extensively studied using bulk experimental methods such as light scattering and rheometry \cite{Watanabe1999,Adam1983,Larson1999}. In recent work, \citet{Wang2017} used small-angle neutron scattering (SANS) to infer molecular relaxation in entangled polymer solutions following a large deformation. Interestingly, these results showed that the fast initial chain retraction step predicted by the D-E model following a step strain was absent from experiments. These findings and recent theoretical advances \cite{Sussman2012,Sussman2012a,Schweizer2016} have brought into question some of the fundamental assumptions of the classic D-E theory and have highlighted the need for new molecular-level studies of entangled polymer solutions \cite{Falzone2015}. Despite their utility in probing polymer dynamics, bulk experimental methods tend to average over large ensembles of molecules, thereby obscuring the role of molecular sub-populations. 
    
    Single molecule techniques allow for the direct observation of polymer chains in flow \cite{Schroeder2018}, thereby revealing dynamic heterogeneity and molecular sub-populations under non-equilibrium conditions. Single molecule fluorescence microscopy (SMFM) has been used to directly observe tube-like or reptative motion in highly entangled DNA solutions \cite{Perkins1994b} and to measure the tube confining potential in entangled DNA solutions \cite{Robertson2007}. Polymer relaxation in unentangled DNA solutions was recently studied using SMFM \cite{Liu2009,Hsiao2017}, and single DNA relaxation in highly entangled solutions following a step strain in shear flow was studied by \citet{Teixeira2007}. Despite recent progress, however, polymer relaxation dynamics in entangled solutions is not fully understood at the molecular level.
    
		 \begin{figure}
 			\includegraphics[scale=1.]{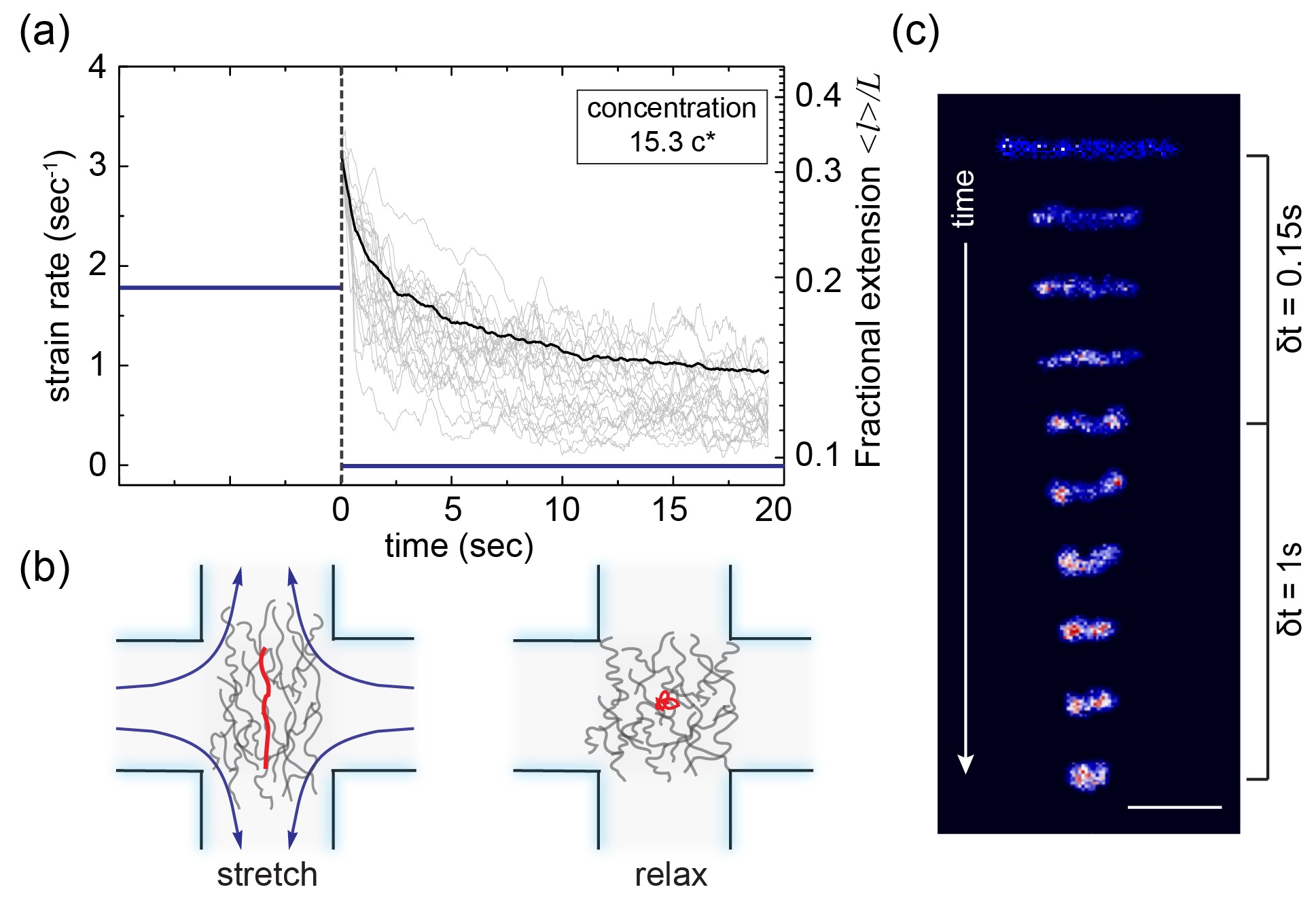}
  			\caption{Single molecule studies of polymer relaxation in entangled DNA solutions. (a) Flow deformation protocol and polymer relaxation process. At time $t$ = 0, the flow is stopped and chains relax to equilibrium. Single molecule trajectories (grey) and ensemble averaged fractional extension (black) at 15.3 $c^*$. (b) Schematic of experiment showing a single fluorescently labeled tracer $\lambda$-DNA molecule (red) in a background of entangled DNA solution. (c) Snapshots of a single tracer DNA molecule relaxing in an entangled solution, showing double-mode relaxation behavior. Scale bar: 5 $\mu$m; $\delta$t is time between images.}
		\end{figure}
	In this letter, we study the relaxation dynamics of single DNA polymers in the cross-over regime between semi-dilute unentangled and entangled solutions using SMFM (Fig. 1). Tracer bacteriophage $\lambda$-DNA molecules (48.5 kbp) are fluorescently labeled with a DNA intercalating dye (YOYO-1) and added to background solutions of unlabeled entangled $\lambda$-DNA (Supporting Information). In this way, we prepared a series of DNA solutions with polymer concentrations between 3.9 $c^*$ and 15.3 $c^*$ (Table S1), where $c^*$=50 $\mu$g/mL is the polymer overlap concentration for $\lambda$-DNA at 22.5 $^\circ$C determined using a combination of dynamic light scattering and Brownian dynamics simulations to account for solvent quality and temperature \cite{Pan2014}. All experiments are conducted in the good solvent regime \cite{Hsiao2017,Schroeder2018} and above the $\theta$-temperature $T_\theta=$ 14.7 $^\circ$C for DNA in aqueous solutions \cite{Pan2014}. Prior work reporting the zero-shear viscosity of monodisperse DNA solutions has shown that $\lambda$-DNA transitions to entangled solution behavior around $c_e \approx$ 3 $c^*$ \cite{Pan2014}, where $c_e$ is the critical entanglement concentration \cite{Graessley1980}. Solutions of $\lambda$-DNA between 3.9 $c^*$ and 15.3 $c^*$ correspond to approximately $n$ $\approx$ 1-12 entanglements per chain (Table S2), as determined by bulk rheology (Supporting Information).
    
    A feedback-controlled microfluidic cross-slot device is used to generate a planar extensional flow (Fig. S1) \cite{Tanyeri2013}. Using this approach, single polymers are stretched to high degrees of extension ($l/L \approx$ 0.6-0.7), where $l$ is the end-to-end polymer extension and $L$ = 21 $\mu$m is the contour length of fluorescently labeled $\lambda$-DNA (Fig. 1a,b). During the deformation step, polymers are exposed to at least $\epsilon$ = $\dot{\epsilon}t$ = 10 units of accumulated fluid strain in extensional flow, and deformation is performed at a dimensionless flow strength called the Weissenberg number $Wi=\dot{\epsilon}\tau_{d} \gg 1$, where $\dot{\epsilon}$ is the strain rate and $\tau_d$ is the reptation time (discussed below). In this way, the flow induces a non-linear deformation prior to relaxation. Following cessation of flow, the relaxation of a single tracer DNA molecule is observed as a function of time (Fig. 1c). Flow field characterization including strain rate determination in entangled DNA solutions is performed via particle tracking velocimetry (PTV), with no elastic instabilities observed under these flow conditions (Fig. S3).

		 \begin{figure*}
 			\includegraphics[scale=1.1]{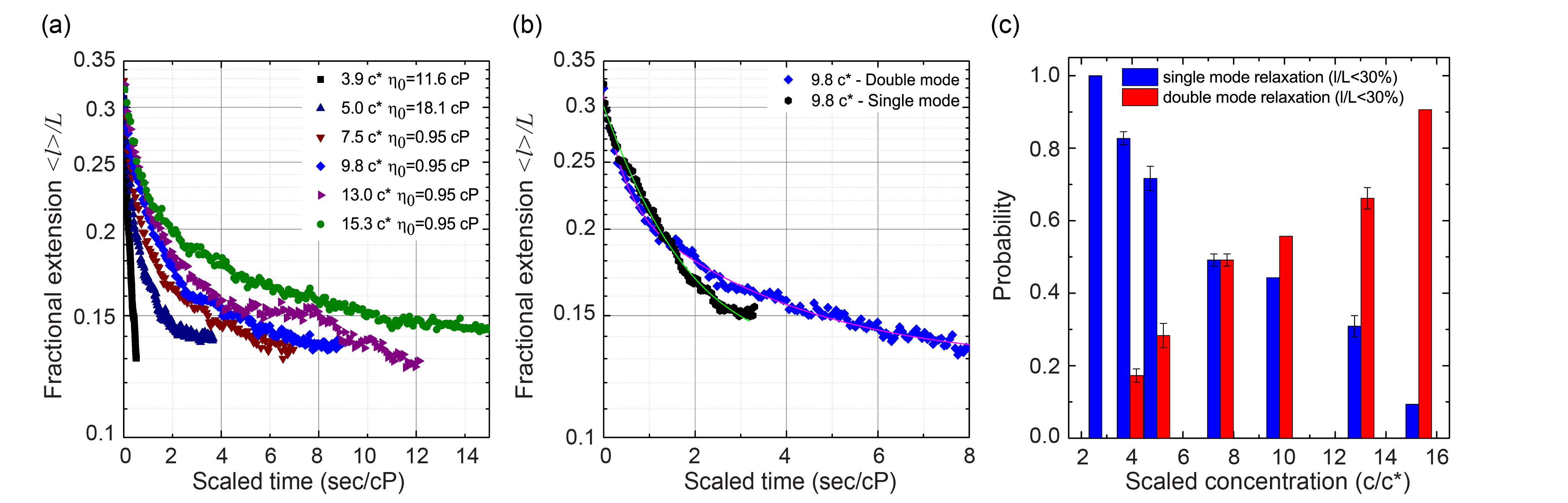}
  			\caption{Single molecule studies of polymer chain relaxation in entangled solutions reveal heterogeneous sub-poplations. (a) Semi-log plot of ensemble-averaged fractional extension $\langle l \rangle/L$ showing relaxation trajectories at five DNA concentrations (5.0 $c^*$, 7.5 $c^*$, 9.8 $c^*$, 13.0 $c^*$, and 15.3 $c^*$; $N\ge 40$ molecules in each ensemble). (b) Molecular sub-populations corresponding to single-mode and double-mode relaxation trajectories for a representative solution concentation at 9.8 c*. Ensemble averaged data for single and double-mode trajectories are shown. (c) Probability distribution of single and double-mode relaxation behavior at different polymer concentrations.}
		\end{figure*}
	
	Polymer relaxation trajectories for the entire molecular ensemble are plotted for different solution concentrations in Fig. 2a. Here, time is scaled by solvent viscosity $\eta_0$ to compare relaxation data between the low concentration 5.0 $c^*$ sample ($\eta_s$ = 18.1 cP) and the remaining solution concentrations ($\eta_s$ = 0.95 cP). Interestingly, for all polymer concentrations, we find that the ensemble-averaged relaxation trajectories cannot be fit by a single-mode exponential decay. Single exponential decay functions are commonly used to analyze polymer extension relaxation data in dilute polymer solutions ($c<c^*$) \cite{Perkins1997a,Zhou2016} and semi-dilute unentangled solutions ($c^*<c<c_e$) \cite{Hsiao2017}. On the other hand, our results show that the underlying molecular ensemble consists of two sub-populations, including polymers that exhibit either a single-mode or a double-mode exponential decay (Fig. 2b and Fig. S4). To classify single polymers into these two different sets, each individual trajectory is fit to both functions, and the best fit is accepted with a suitable adjusted R-square value $\ge$90$\%$ (Fig. S5). The single-mode relaxation time $\tau_s$ is determined by fitting the terminal 30$\%$ of the squared polymer extension $(l/L)^2$ to a single-mode exponential decay:
	\begin{equation}
	(l/L)^2 = A\exp(-t/\tau_s) + B
	\end{equation}
where $A$ and $B$ are numerical constants. The double-mode relaxation times $\tau_{d,1}$ and $\tau_{d,2}$ are obtained by fitting $(l/L)^2$ to a double-mode exponential decay: 
	\begin{equation}
	(l/L)^2 = A_1 \exp(-t/\tau_{d,1}) + A_2 \exp(-t/\tau_{d,2}) + B
	\end{equation}
where $A_1$, $A_2$, and $B$ are numerical constants.	 

	Molecular ensembles corresponding to single-mode and double-mode relaxation behavior are shown in Fig. 2b and Fig. S4. Polymers exhibiting double-mode relaxation behavior exhibit an initially fast retraction with a characteristic timescale $\tau_{d,1}$, followed by slower relaxation with timescale $\tau_{d,2}$ before returning an equilibrium coiled state. A histogram showing the probability of single-mode and double-mode relaxation behavior as a function of polymer concentration is shown in Fig. 2c. Upon increasing polymer concentration concentration from 2.8 $c^*$ to 15.3 $c^*$, the probability of single-mode relaxation behavior decreases, whereas the probability of double-mode behavior increases. The emergence of multiple molecular sub-populations is consistent with the gradual transition from the semi-dilute unentangled regime ($c<c_e$) to the semi-dilute entangled regime ($c>c_e$) at a critical entanglement concentration $c_e\approx$ 3 $c^*$. This value of $c_e$ is consistent with prior work from bulk shear rheology of DNA \cite{Pan2014} and single molecule measurements of polymer diffusion \cite{Robertson2007a}. At relatively high polymer concentrations ($c$ = 15.3 $c^*$ $\approx$ 5.1 $c_e$), nearly all relaxation trajectories exhibit double-mode relaxation behavior, which is consistent with prior observations on highly entangled $\lambda$-DNA solutions \cite{Teixeira2007}.
    
	We quantitatively determined the single-mode relaxation times $\tau_s$ and double-mode relaxation times $\tau_{d,1}$ and $\tau_{d,2}$ from our experiments (Table S2). In this way, we observe clear power-law scaling behavior for the longest relaxation times as a function of scaled concentration $c/c^*$ across a wide range of polymer concentrations, as shown in Fig. 3. Results from single molecule experiments are compared to longest relaxation times of entangled $\lambda$-DNA solutions measured from bulk shear rheology (based on a relaxation time $\lambda$ from zero-shear viscosity $\eta_0$) \cite{Pan2014}, single molecule experiments following cessation of shear flow \cite{Teixeira2007}, and single polymer diffusion measurements \cite{Gong2014}. For bulk experiments and single molecule measurements, the relaxation times $\lambda$ and $\tau$ are normalized by the longest polymer relaxation time in the dilute limit $\lambda_z$ and $\tau_z$, respectively, and plotted as a function of scaled concentration $c/c^*$ in Fig. 3. 
	
		 \begin{figure}
 			\includegraphics[scale=1.3]{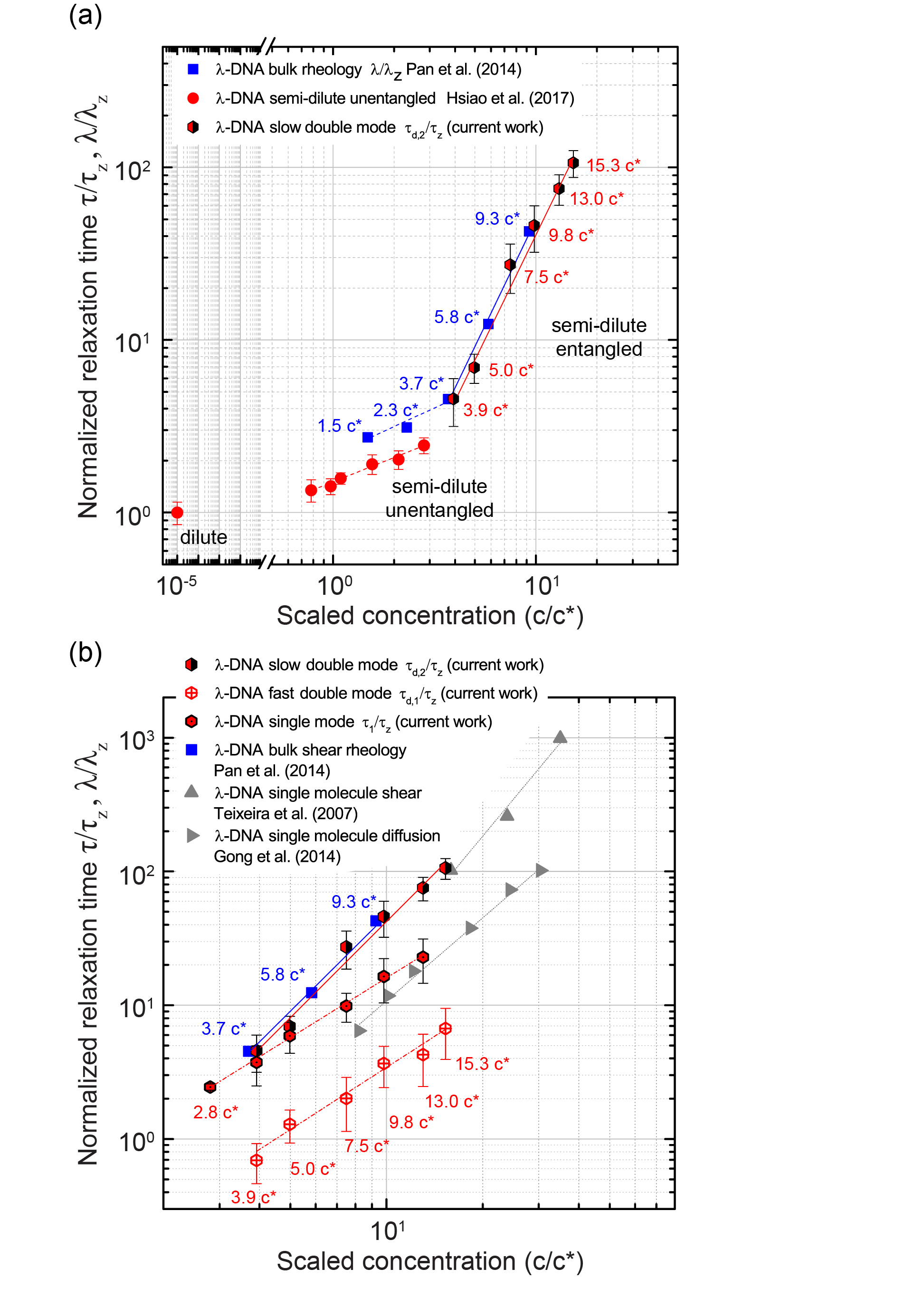}
  			\caption{Normalized longest relaxation times $\tau/\tau_z$ and $\lambda/\lambda_z$ as a function of scaled polymer concentration $c/c^*$. (a) Power-law relaxation time scaling behavior across the semi-dilute unentangled and entangled regimes. (b) Characteristic longest relaxation times in the entangled regime.}
		\end{figure}
	
	Figure 3a shows the concentration-dependent power-law scalings of the longest relaxation times across the semi-dilute unentangled ($c^* < c < c_e$) and entangled regime ($c > c_e$). In the semi-dilute unentangled regime, the longest relaxation time scales with polymer concentration as $\tau/\tau_z \sim (c/c^*)^{0.48}$, as previously reported \cite{Hsiao2017}. In the entangled regime, the relaxation behavior shows a dramatic slow down in dynamics. Here, the slower double-mode timescale $\tau_{d,2}$ exhibits a power-law scaling consistent with the characteristic reptation time for semi-dilute entangled polymer solutions \cite{DeGennes1976b,Rubinstein2003}. In particular, we find $\tau_{d,2}/\tau_z \sim (c/c^*)^{2.4}$ from single molecule experiments, which compares favorably with relaxation time scalings from bulk shear rheological experiments $\lambda/\lambda_z \sim (c/c^*)^{2.4}$ \cite{Pan2014}. In entangled polymer melts, experiments \cite{Lodge1999} show that the reptation time $\tau_d$ exhibits a power-law scaling with polymer molecular weight $M$ such that:
	\begin{equation}
		\tau_d \sim M^{3.4}
	\end{equation}  
On the other hand, in entangled polymer solutions, polymer concentration and solvent quality both play a role on the reptation time $\tau_d$. Scaling theory can be used to derive the concentration and solvent quality dependence of $\tau_d$ \cite{Rubinstein2003} (Supporting Information), such that:
	\begin{equation}
	\tau_d =  \frac{\tau_z}{(N_e(1))^{1.4}} \left( \frac{c}{c^*} \right)^{\frac{3.4-3\nu}{3\nu-1}} (zN^{-0.5})^{6\nu-5.8} (z^{1.4})^{2\nu+1}
		\end{equation}	
where $\tau_z$ is the polymer relaxation time in the dilute limit, $\nu$ is the effective excluded volume coefficient, $N$ is the number of statistical steps in the polymer (Kuhn segments), and $N_e(1)$ is the number of Kuhn steps in one entanglement strand in a melt. Moreover, $z$ is the chain interaction parameter which is a measure of solvent quality (Supporting Information) \cite{Rubinstein2003}. Briefly, $z=k\left( 1 - T_{\theta} / T \right) \sqrt{M}$, where $M$ is polymer molecular weight and the constant $k$ has been determined for DNA solutions using a combination of light scattering and BD simulations \cite{Pan2014}, thereby enabling calculation of $z$ for any $M$ and $T$. For our experiments on $\lambda$-DNA conducted at $T$ = 22.5 $^o$C, we find $z \approx$ 0.71, which corresponds to the lower limit of the good solvent regime \cite{Samsal2017}. 

	Given that the reptation time $\tau_d$ scales with concentration as $\tau_d \sim (c/c^*)^{(3.4-3\nu)/(3\nu-1)}$ for entangled solutions and $\tau_{d,2} \sim (c/c^*)^{2.4}$ from our data, we determined an effective excluded volume exponent $\nu \approx 0.57$, which is consistent with good solvent conditions. Fig. 3b also shows prior single molecule experimental data from \citet{Teixeira2007}, where double-mode relaxation behavior following shear flow deformation was observed at high DNA concentrations (16 $c^*$ - 35 $c^*$). Analysis of these prior data shows $\tau_{d,2} \sim (c/c^*)^{2.9}$, which is a steeper concentration dependence than the current work and corresponds to an effective excluded volume exponent of $\nu \approx$ 0.53, which is closer to the expected scalings for $\Theta$-solvent conditions \cite{Rubinstein2003, Liu2009}. Interestingly, the experiments of \citet{Teixeira2007} were performed at $T$ = 18 $^o$C, such that the chain interaction parameter $z \approx$ 0.3, suggesting near $\Theta$-solvent conditions. Together, these data show the sensitivity of polymer relaxation behavior to experimental conditions for entangled polymer solutions.
    
    The fast double-mode relaxation time $\tau_{d,1}$ exhibits a weaker power-law concentration dependence compared to the slow double-mode time $\tau_{d,2}$, such that $\tau_{d,1}/\tau_z \sim (c/c^*)^{1.5}$ (Fig. 3b). Interestingly, the numerical values of $\tau_{d,1}$ are on the order of the Rouse time $\tau_R = 6R_G^2 / 3 \pi^2 D_G^2$ \cite{Doi1986} for $\lambda$-DNA (Table S2), where $R_G$ is radius of gyration determined from universal scaling relations for DNA \cite{Pan2014} and $D_G$ is the center-of-mass diffusion coefficient determined in prior single molecule experiments \cite{Robertson2006a}. In this way, we determined a Rouse time $\tau_R$ = 0.3 sec for $\lambda$-DNA at $T$ = 22.5$^o$C in a solvent viscosity $\eta_s$ = 1.0 cP, which is consistent with prior estimates of $\tau_R$ \cite{Robertson2007}. From this view, we hypothesize that $\tau_{d,1}$ corresponds to a timescale associated with a Rouse-like chain recovery or chain retraction following the non-linear chain stretching step. A similar fast initial stress decay following the cessation of a large uniaxial extensional deformation has been observed for entangled polymer solutions in bulk experiments \cite{Bhattacharjee2003}.
    
    However, despite the apparent similarity to Rouse-like chain behavior, our data shows that the fast double-mode relaxation time $\tau_{d,1}$ is concentration dependent, unlike a true Rouse-like response. These results suggest that the initial fast retraction step slows down as the local polymer concentration increases, thereby increasing chain friction due to nearby polymer chains. At longer times, entanglements reform in this molecular sub-population, and the polymer chain transitions to a reptative relaxation process described by $\tau_{d,2}$.
    
	Single molecule experiments further reveal an additional relaxation time $\tau_s$, which emerges through a different molecular sub-population in the ensemble. The single-mode relaxation time exhibits a power-law concentration scaling such that $\tau_s / \tau_z \sim (c/c^*)^{1.5}$. Although $\tau_{d,1}$ and $\tau_s$ show nearly the same power-law scaling with concentration, $\tau_{s}$ is approximately a factor of 5 larger than $\tau_{d,1}$ (Table S2), which suggests a different physical origin for $\tau_{s}$ compared to $\tau_{d,1}$. We conjecture that $\tau_s$ corresponds to the timescale of a polymer chain relaxation in a locally unentangled environment in the polymer solution. For this molecular sub-population, polymer chains experience no chain-chain entanglements during the relaxation event, yet they may experience enhanced intermolecular interactions with an associated increase in chain friction during the relaxation process. Interestingly, locally unentangled behavior only exists in the transition regime from semi-dilute unentangled to semi-dilute entangled solutions (Fig. 2c). At high polymer concentrations ($c \ge 16$ $c^*$), the single-mode relaxation behavior is absent. These results are further supported by bulk shear rheology measurements on our entangled DNA solutions, where an entanglement plateau emerges at 5.0 $c^*$ and is clearly observed around 15 $c^*$ (Fig. S6).
    
    Our single molecule experiments reveal several intriguing features of polymer chain relaxation following a non-linear deformation. In the transition regime from unentangled to entangled polymer solutions, these results reveal two distinct molecular sub-populations exhibiting single-mode and double-mode exponential relaxation behavior. We conjecture that these two relaxation modes correspond to different molecular relaxation pathways, such that the slow double-mode timescale $\tau_{d,2}$ is attributed to slow relaxation dynamics associated with polymer reptation. The fast double-mode relaxation time $\tau_{d,1}$ is attributed to the fast initial chain retraction step immediately following deformation. On the other hand, we hypothesize that the single-mode time $\tau_s$ emerges from local regions of the polymer solution that have become transiently unentangled due to the strong deformation. These results suggest that an apparently well-mixed polymer solution may be entangled at thermal equilibrium and can become transiently disentangled upon exposure to strong deformation. As polymer concentration is increased ($c \gg c^*$), the propensity for transient disentanglements to occur within the solution decreases. Taken together, our work provides new molecular-level perspectives on entangled polymer solutions that are facilitated by single molecule observations.
		
\begin{acknowledgments}
We thank Kenneth Schweizer for helpful discussions and Simon Rogers and Johnny Lee for rheometeric measurements. This work was supported by NSF CBET 1604038 for C.M.S. and a PPG-MRL graduate research assistantship award for Y.Z.
\end{acknowledgments}




\begin{thebibliography}{38}%
\makeatletter
\providecommand \@ifxundefined [1]{%
 \@ifx{#1\undefined}
}%
\providecommand \@ifnum [1]{%
 \ifnum #1\expandafter \@firstoftwo
 \else \expandafter \@secondoftwo
 \fi
}%
\providecommand \@ifx [1]{%
 \ifx #1\expandafter \@firstoftwo
 \else \expandafter \@secondoftwo
 \fi
}%
\providecommand \natexlab [1]{#1}%
\providecommand \enquote  [1]{``#1''}%
\providecommand \bibnamefont  [1]{#1}%
\providecommand \bibfnamefont [1]{#1}%
\providecommand \citenamefont [1]{#1}%
\providecommand \href@noop [0]{\@secondoftwo}%
\providecommand \href [0]{\begingroup \@sanitize@url \@href}%
\providecommand \@href[1]{\@@startlink{#1}\@@href}%
\providecommand \@@href[1]{\endgroup#1\@@endlink}%
\providecommand \@sanitize@url [0]{\catcode `\\12\catcode `\$12\catcode
  `\&12\catcode `\#12\catcode `\^12\catcode `\_12\catcode `\%12\relax}%
\providecommand \@@startlink[1]{}%
\providecommand \@@endlink[0]{}%
\providecommand \url  [0]{\begingroup\@sanitize@url \@url }%
\providecommand \@url [1]{\endgroup\@href {#1}{\urlprefix }}%
\providecommand \urlprefix  [0]{URL }%
\providecommand \Eprint [0]{\href }%
\providecommand \doibase [0]{http://dx.doi.org/}%
\providecommand \selectlanguage [0]{\@gobble}%
\providecommand \bibinfo  [0]{\@secondoftwo}%
\providecommand \bibfield  [0]{\@secondoftwo}%
\providecommand \translation [1]{[#1]}%
\providecommand \BibitemOpen [0]{}%
\providecommand \bibitemStop [0]{}%
\providecommand \bibitemNoStop [0]{.\EOS\space}%
\providecommand \EOS [0]{\spacefactor3000\relax}%
\providecommand \BibitemShut  [1]{\csname bibitem#1\endcsname}%
\let\auto@bib@innerbib\@empty
\bibitem [{\citenamefont {{De Gennes}}(1976{\natexlab{a}})}]{DeGennes1976a}%
  \BibitemOpen
  \bibfield  {author} {\bibinfo {author} {\bibfnamefont {P.~G.}\ \bibnamefont
  {{De Gennes}}},\ }\href {\doibase 10.1021/ma60052a011} {\bibfield  {journal}
  {\bibinfo  {journal} {Macromolecules}\ }\textbf {\bibinfo {volume} {9}},\
  \bibinfo {pages} {587} (\bibinfo {year} {1976}{\natexlab{a}})}\BibitemShut
  {NoStop}%
\bibitem [{\citenamefont {Doi}\ and\ \citenamefont {Edwards}(1986)}]{Doi1986}%
  \BibitemOpen
  \bibfield  {author} {\bibinfo {author} {\bibfnamefont {M.}~\bibnamefont
  {Doi}}\ and\ \bibinfo {author} {\bibfnamefont {S.~F.}\ \bibnamefont
  {Edwards}},\ }\href@noop {} {\emph {\bibinfo {title} {{The Theory of Polymer
  Dynamics}}}}\ (\bibinfo  {publisher} {Oxford University Press},\ \bibinfo
  {address} {Oxford, UK},\ \bibinfo {year} {1986})\ p.\ \bibinfo {pages}
  {391}\BibitemShut {NoStop}%
\bibitem [{\citenamefont {{De Gennes}}(1976{\natexlab{b}})}]{DeGennes1976b}%
  \BibitemOpen
  \bibfield  {author} {\bibinfo {author} {\bibfnamefont {P.~G.}\ \bibnamefont
  {{De Gennes}}},\ }\href@noop {} {\bibfield  {journal} {\bibinfo  {journal}
  {Macomolecules}\ }\textbf {\bibinfo {volume} {9}},\ \bibinfo {pages} {594}
  (\bibinfo {year} {1976}{\natexlab{b}})}\BibitemShut {NoStop}%
\bibitem [{\citenamefont {Rubinstein}\ and\ \citenamefont
  {Colby}(2003)}]{Rubinstein2003}%
  \BibitemOpen
  \bibfield  {author} {\bibinfo {author} {\bibfnamefont {M.}~\bibnamefont
  {Rubinstein}}\ and\ \bibinfo {author} {\bibfnamefont {R.~H.}\ \bibnamefont
  {Colby}},\ }\href@noop {} {\emph {\bibinfo {title} {{Polymer Physics}}}}\
  (\bibinfo  {publisher} {Oxford University Press},\ \bibinfo {address}
  {Oxford, UK},\ \bibinfo {year} {2003})\ p.\ \bibinfo {pages}
  {440}\BibitemShut {NoStop}%
\bibitem [{\citenamefont {Schweizer}\ and\ \citenamefont
  {Sussman}(2016)}]{Schweizer2016}%
  \BibitemOpen
  \bibfield  {author} {\bibinfo {author} {\bibfnamefont {K.~S.}\ \bibnamefont
  {Schweizer}}\ and\ \bibinfo {author} {\bibfnamefont {D.~M.}\ \bibnamefont
  {Sussman}},\ }\href {\doibase 10.1063/1.4968516} {\bibfield  {journal}
  {\bibinfo  {journal} {The Journal of Chemical Physics}\ }\textbf {\bibinfo
  {volume} {145}},\ \bibinfo {pages} {214903} (\bibinfo {year}
  {2016})}\BibitemShut {NoStop}%
\bibitem [{\citenamefont {Lodge}\ \emph {et~al.}(1990)\citenamefont {Lodge},
  \citenamefont {Rotstein},\ and\ \citenamefont {Prager}}]{Lodge1990}%
  \BibitemOpen
  \bibfield  {author} {\bibinfo {author} {\bibfnamefont {T.~P.}\ \bibnamefont
  {Lodge}}, \bibinfo {author} {\bibfnamefont {N.~A.}\ \bibnamefont {Rotstein}},
  \ and\ \bibinfo {author} {\bibfnamefont {S.}~\bibnamefont {Prager}},\ }in\
  \href
  {https://books.google.com/books?hl=en{\&}lr={\&}id=bc3XBck5w2gC{\&}oi=fnd{\&}pg=PA1{\&}dq=T.+P.+Lodge,+N.+A.+Rotstein,+and+S.+Prager,+in+Advances+in+Chemical+Physics,+Vol.+79,+{\&}ots=BAfT8LZz9F{\&}sig=3d8D6dKJzhGBQXUmzdxBMhirn8w{\#}v=onepage{\&}q{\&}f=false}
  {\emph {\bibinfo {booktitle} {Advances in Chemical Physics, Volume 79}}},\
  \bibinfo {editor} {edited by\ \bibinfo {editor} {\bibfnamefont
  {I.}~\bibnamefont {Prigogine}}\ and\ \bibinfo {editor} {\bibfnamefont
  {S.~A.}\ \bibnamefont {Rice}}}\ (\bibinfo  {publisher} {John Wiley {\&}
  Sons},\ \bibinfo {year} {1990})\ pp.\ \bibinfo {pages} {1--132}\BibitemShut
  {NoStop}%
\bibitem [{\citenamefont {McLeish}(2002)}]{McLeish2002a}%
  \BibitemOpen
  \bibfield  {author} {\bibinfo {author} {\bibfnamefont {T.~C.~B.}\
  \bibnamefont {McLeish}},\ }\href {\doibase 10.1080/00018730210153216}
  {\bibfield  {journal} {\bibinfo  {journal} {Advances in Physics}\ }\textbf
  {\bibinfo {volume} {51}},\ \bibinfo {pages} {1379} (\bibinfo {year}
  {2002})}\BibitemShut {NoStop}%
\bibitem [{\citenamefont {Ferry}(1980)}]{Ferry1980}%
  \BibitemOpen
  \bibfield  {author} {\bibinfo {author} {\bibfnamefont {J.~D.}\ \bibnamefont
  {Ferry}},\ }\href
  {https://books.google.com/books?hl=en{\&}lr={\&}id=9dqQY3Ujsx4C{\&}pgis=1}
  {\emph {\bibinfo {title} {{Viscoelastic Properties of Polymers}}}},\ \bibinfo
  {edition} {3rd}\ ed.\ (\bibinfo  {publisher} {John Wiley {\&} Sons},\
  \bibinfo {address} {New York, NY, USA},\ \bibinfo {year} {1980})\BibitemShut
  {NoStop}%
\bibitem [{\citenamefont {Lodge}(1999)}]{Lodge1999}%
  \BibitemOpen
  \bibfield  {author} {\bibinfo {author} {\bibfnamefont {T.~P.}\ \bibnamefont
  {Lodge}},\ }\href {\doibase 10.1103/PhysRevLett.83.3218} {\bibfield
  {journal} {\bibinfo  {journal} {Physical Review Letters}\ }\textbf {\bibinfo
  {volume} {83}},\ \bibinfo {pages} {3218} (\bibinfo {year}
  {1999})}\BibitemShut {NoStop}%
\bibitem [{\citenamefont {Larson}(1999)}]{Larson1999}%
  \BibitemOpen
  \bibfield  {author} {\bibinfo {author} {\bibfnamefont {R.~G.}\ \bibnamefont
  {Larson}},\ }\href@noop {} {\emph {\bibinfo {title} {{The Structure and
  Rheology of Complex Fluids}}}}\ (\bibinfo  {publisher} {Oxford University
  Press},\ \bibinfo {address} {New York, NY, USA},\ \bibinfo {year}
  {1999})\BibitemShut {NoStop}%
\bibitem [{\citenamefont {Graessley}(1982)}]{Graessley1982}%
  \BibitemOpen
  \bibfield  {author} {\bibinfo {author} {\bibfnamefont {W.}~\bibnamefont
  {Graessley}},\ }in\ \href {\doibase doi: 10.1007/bfb0038532} {\emph {\bibinfo
  {booktitle} {Synthesis and Degradation Rheology and Extrusion}}}\ (\bibinfo
  {publisher} {Springer},\ \bibinfo {year} {1982})\ pp.\ \bibinfo {pages}
  {67--117}\BibitemShut {NoStop}%
\bibitem [{\citenamefont {Viovy}\ \emph {et~al.}(1991)\citenamefont {Viovy},
  \citenamefont {Rubinstein},\ and\ \citenamefont {Colby}}]{Viovy1991}%
  \BibitemOpen
  \bibfield  {author} {\bibinfo {author} {\bibfnamefont {J.~L.}\ \bibnamefont
  {Viovy}}, \bibinfo {author} {\bibfnamefont {M.}~\bibnamefont {Rubinstein}}, \
  and\ \bibinfo {author} {\bibfnamefont {R.~H.}\ \bibnamefont {Colby}},\ }\href
  {\doibase 10.1021/ma00012a020} {\bibfield  {journal} {\bibinfo  {journal}
  {Macromolecules}\ }\textbf {\bibinfo {volume} {24}},\ \bibinfo {pages} {3587}
  (\bibinfo {year} {1991})}\BibitemShut {NoStop}%
\bibitem [{\citenamefont {Doi}(1983)}]{Doi1983}%
  \BibitemOpen
  \bibfield  {author} {\bibinfo {author} {\bibfnamefont {M.}~\bibnamefont
  {Doi}},\ }\href {\doibase 10.1002/pol.1983.180210501} {\bibfield  {journal}
  {\bibinfo  {journal} {Journal of Polymer Science: Polymer Physics Edition}\
  }\textbf {\bibinfo {volume} {21}},\ \bibinfo {pages} {667} (\bibinfo {year}
  {1983})}\BibitemShut {NoStop}%
\bibitem [{\citenamefont {Milner}\ and\ \citenamefont
  {McLeish}(1998)}]{Milner1998}%
  \BibitemOpen
  \bibfield  {author} {\bibinfo {author} {\bibfnamefont {S.}~\bibnamefont
  {Milner}}\ and\ \bibinfo {author} {\bibfnamefont {T.}~\bibnamefont
  {McLeish}},\ }\href {\doibase 10.1103/PhysRevLett.81.725} {\bibfield
  {journal} {\bibinfo  {journal} {Physical Review Letters}\ }\textbf {\bibinfo
  {volume} {81}},\ \bibinfo {pages} {725} (\bibinfo {year} {1998})}\BibitemShut
  {NoStop}%
\bibitem [{\citenamefont {Marrucci}\ and\ \citenamefont
  {Grizzuti}(1988)}]{Marrucci1988}%
  \BibitemOpen
  \bibfield  {author} {\bibinfo {author} {\bibfnamefont {G.}~\bibnamefont
  {Marrucci}}\ and\ \bibinfo {author} {\bibfnamefont {N.}~\bibnamefont
  {Grizzuti}},\ }\href@noop {} {\bibfield  {journal} {\bibinfo  {journal}
  {Gazzetta Chimica Italiana}\ }\textbf {\bibinfo {volume} {118}},\ \bibinfo
  {pages} {179} (\bibinfo {year} {1988})}\BibitemShut {NoStop}%
\bibitem [{\citenamefont {Graham}\ \emph {et~al.}(2003)\citenamefont {Graham},
  \citenamefont {Likhtman}, \citenamefont {McLeish},\ and\ \citenamefont
  {Milner}}]{Graham2003}%
  \BibitemOpen
  \bibfield  {author} {\bibinfo {author} {\bibfnamefont {R.~S.}\ \bibnamefont
  {Graham}}, \bibinfo {author} {\bibfnamefont {A.~E.}\ \bibnamefont
  {Likhtman}}, \bibinfo {author} {\bibfnamefont {T.~C.~B.}\ \bibnamefont
  {McLeish}}, \ and\ \bibinfo {author} {\bibfnamefont {S.~T.}\ \bibnamefont
  {Milner}},\ }\href {\doibase 10.1122/1.1595099} {\bibfield  {journal}
  {\bibinfo  {journal} {Journal of Rheology}\ }\textbf {\bibinfo {volume}
  {47}},\ \bibinfo {pages} {1171} (\bibinfo {year} {2003})}\BibitemShut
  {NoStop}%
\bibitem [{\citenamefont {Watanabe}(1999)}]{Watanabe1999}%
  \BibitemOpen
  \bibfield  {author} {\bibinfo {author} {\bibfnamefont {H.}~\bibnamefont
  {Watanabe}},\ }\href {\doibase 10.1016/S0079-6700(99)00029-5} {\bibfield
  {journal} {\bibinfo  {journal} {Progress in Polymer Science}\ }\textbf
  {\bibinfo {volume} {24}},\ \bibinfo {pages} {1253} (\bibinfo {year}
  {1999})}\BibitemShut {NoStop}%
\bibitem [{\citenamefont {Adam}\ and\ \citenamefont
  {Delsanti}(1983)}]{Adam1983}%
  \BibitemOpen
  \bibfield  {author} {\bibinfo {author} {\bibfnamefont {M.}~\bibnamefont
  {Adam}}\ and\ \bibinfo {author} {\bibfnamefont {M.}~\bibnamefont
  {Delsanti}},\ }\href {\doibase 10.1051/jphys:019840045090151300} {\bibfield
  {journal} {\bibinfo  {journal} {Journal de Physique}\ }\textbf {\bibinfo
  {volume} {43}},\ \bibinfo {pages} {1185} (\bibinfo {year}
  {1983})}\BibitemShut {NoStop}%
\bibitem [{\citenamefont {Wang}\ \emph {et~al.}(2017)\citenamefont {Wang},
  \citenamefont {Lam}, \citenamefont {Chen}, \citenamefont {Wang},
  \citenamefont {Liu}, \citenamefont {Liu}, \citenamefont {Porcar},
  \citenamefont {Stanley}, \citenamefont {Zhao}, \citenamefont {Hong},\ and\
  \citenamefont {Wang}}]{Wang2017}%
  \BibitemOpen
  \bibfield  {author} {\bibinfo {author} {\bibfnamefont {Z.}~\bibnamefont
  {Wang}}, \bibinfo {author} {\bibfnamefont {C.~N.}\ \bibnamefont {Lam}},
  \bibinfo {author} {\bibfnamefont {W.-R.}\ \bibnamefont {Chen}}, \bibinfo
  {author} {\bibfnamefont {W.}~\bibnamefont {Wang}}, \bibinfo {author}
  {\bibfnamefont {J.}~\bibnamefont {Liu}}, \bibinfo {author} {\bibfnamefont
  {Y.}~\bibnamefont {Liu}}, \bibinfo {author} {\bibfnamefont {L.}~\bibnamefont
  {Porcar}}, \bibinfo {author} {\bibfnamefont {C.~B.}\ \bibnamefont {Stanley}},
  \bibinfo {author} {\bibfnamefont {Z.}~\bibnamefont {Zhao}}, \bibinfo {author}
  {\bibfnamefont {K.}~\bibnamefont {Hong}}, \ and\ \bibinfo {author}
  {\bibfnamefont {Y.}~\bibnamefont {Wang}},\ }\href {\doibase
  10.1103/PhysRevX.7.031003} {\bibfield  {journal} {\bibinfo  {journal}
  {Physical Review X}\ }\textbf {\bibinfo {volume} {7}},\ \bibinfo {pages}
  {031003} (\bibinfo {year} {2017})}\BibitemShut {NoStop}%
\bibitem [{\citenamefont {Sussman}\ and\ \citenamefont
  {Schweizer}(2012{\natexlab{a}})}]{Sussman2012}%
  \BibitemOpen
  \bibfield  {author} {\bibinfo {author} {\bibfnamefont {D.~M.}\ \bibnamefont
  {Sussman}}\ and\ \bibinfo {author} {\bibfnamefont {K.~S.}\ \bibnamefont
  {Schweizer}},\ }\href {\doibase 10.1021/ma300006s} {\bibfield  {journal}
  {\bibinfo  {journal} {Macromolecules}\ }\textbf {\bibinfo {volume} {45}},\
  \bibinfo {pages} {3270} (\bibinfo {year} {2012}{\natexlab{a}})}\BibitemShut
  {NoStop}%
\bibitem [{\citenamefont {Sussman}\ and\ \citenamefont
  {Schweizer}(2012{\natexlab{b}})}]{Sussman2012a}%
  \BibitemOpen
  \bibfield  {author} {\bibinfo {author} {\bibfnamefont {D.~M.}\ \bibnamefont
  {Sussman}}\ and\ \bibinfo {author} {\bibfnamefont {K.~S.}\ \bibnamefont
  {Schweizer}},\ }\href {\doibase 10.1103/PhysRevLett.109.168306} {\bibfield
  {journal} {\bibinfo  {journal} {Physical Review Letters}\ }\textbf {\bibinfo
  {volume} {109}},\ \bibinfo {pages} {1} (\bibinfo {year}
  {2012}{\natexlab{b}})}\BibitemShut {NoStop}%
\bibitem [{\citenamefont {Falzone}\ and\ \citenamefont
  {Robertson-Anderson}(2015)}]{Falzone2015}%
  \BibitemOpen
  \bibfield  {author} {\bibinfo {author} {\bibfnamefont {T.~T.}\ \bibnamefont
  {Falzone}}\ and\ \bibinfo {author} {\bibfnamefont {R.~M.}\ \bibnamefont
  {Robertson-Anderson}},\ }\href {\doibase 10.1021/acsmacrolett.5b00673}
  {\bibfield  {journal} {\bibinfo  {journal} {ACS Macro Letters}\ }\textbf
  {\bibinfo {volume} {4}},\ \bibinfo {pages} {1194} (\bibinfo {year}
  {2015})}\BibitemShut {NoStop}%
\bibitem [{\citenamefont {Schroeder}(2018)}]{Schroeder2018}%
  \BibitemOpen
  \bibfield  {author} {\bibinfo {author} {\bibfnamefont {C.~M.}\ \bibnamefont
  {Schroeder}},\ }\href {\doibase 10.1122/1.5013246} {\bibfield  {journal}
  {\bibinfo  {journal} {Journal of Rheology}\ }\textbf {\bibinfo {volume}
  {62}},\ \bibinfo {pages} {371} (\bibinfo {year} {2018})}\BibitemShut
  {NoStop}%
\bibitem [{\citenamefont {Perkins}\ \emph {et~al.}(1994)\citenamefont
  {Perkins}, \citenamefont {Smith},\ and\ \citenamefont {Chu}}]{Perkins1994b}%
  \BibitemOpen
  \bibfield  {author} {\bibinfo {author} {\bibfnamefont {T.~T.}\ \bibnamefont
  {Perkins}}, \bibinfo {author} {\bibfnamefont {D.~E.}\ \bibnamefont {Smith}},
  \ and\ \bibinfo {author} {\bibfnamefont {S.}~\bibnamefont {Chu}},\ }\href
  {\doibase 10.1126/science.8171335} {\bibfield  {journal} {\bibinfo  {journal}
  {Science}\ }\textbf {\bibinfo {volume} {264}},\ \bibinfo {pages} {819}
  (\bibinfo {year} {1994})}\BibitemShut {NoStop}%
\bibitem [{\citenamefont {Robertson}\ and\ \citenamefont
  {Smith}(2007{\natexlab{a}})}]{Robertson2007}%
  \BibitemOpen
  \bibfield  {author} {\bibinfo {author} {\bibfnamefont {R.~M.}\ \bibnamefont
  {Robertson}}\ and\ \bibinfo {author} {\bibfnamefont {D.~E.}\ \bibnamefont
  {Smith}},\ }\href {\doibase 10.1103/PhysRevLett.99.126001} {\bibfield
  {journal} {\bibinfo  {journal} {Physical Review Letters}\ }\textbf {\bibinfo
  {volume} {99}},\ \bibinfo {pages} {8} (\bibinfo {year}
  {2007}{\natexlab{a}})}\BibitemShut {NoStop}%
\bibitem [{\citenamefont {Liu}\ \emph {et~al.}(2009)\citenamefont {Liu},
  \citenamefont {Jun},\ and\ \citenamefont {Steinberg}}]{Liu2009}%
  \BibitemOpen
  \bibfield  {author} {\bibinfo {author} {\bibfnamefont {Y.}~\bibnamefont
  {Liu}}, \bibinfo {author} {\bibfnamefont {Y.}~\bibnamefont {Jun}}, \ and\
  \bibinfo {author} {\bibfnamefont {V.}~\bibnamefont {Steinberg}},\ }\href
  {\doibase 10.1122/1.3160734} {\bibfield  {journal} {\bibinfo  {journal}
  {Journal of Rheology}\ }\textbf {\bibinfo {volume} {53}},\ \bibinfo {pages}
  {1069} (\bibinfo {year} {2009})}\BibitemShut {NoStop}%
\bibitem [{\citenamefont {Hsiao}\ \emph {et~al.}(2017)\citenamefont {Hsiao},
  \citenamefont {Samsal}, \citenamefont {Prakash},\ and\ \citenamefont
  {Schroeder}}]{Hsiao2017}%
  \BibitemOpen
  \bibfield  {author} {\bibinfo {author} {\bibfnamefont {K.-W.}\ \bibnamefont
  {Hsiao}}, \bibinfo {author} {\bibfnamefont {C.}~\bibnamefont {Samsal}},
  \bibinfo {author} {\bibfnamefont {J.~R.}\ \bibnamefont {Prakash}}, \ and\
  \bibinfo {author} {\bibfnamefont {C.~M.}\ \bibnamefont {Schroeder}},\ }\href
  {\doibase 10.1122/1.4972236} {\bibfield  {journal} {\bibinfo  {journal}
  {Journal of Rheology}\ }\textbf {\bibinfo {volume} {61}},\ \bibinfo {pages}
  {151} (\bibinfo {year} {2017})}\BibitemShut {NoStop}%
\bibitem [{\citenamefont {Teixeira}\ \emph {et~al.}(2007)\citenamefont
  {Teixeira}, \citenamefont {Dambal}, \citenamefont {Richter}, \citenamefont
  {Shaqfeh},\ and\ \citenamefont {Chu}}]{Teixeira2007}%
  \BibitemOpen
  \bibfield  {author} {\bibinfo {author} {\bibfnamefont {R.~E.}\ \bibnamefont
  {Teixeira}}, \bibinfo {author} {\bibfnamefont {A.~K.}\ \bibnamefont
  {Dambal}}, \bibinfo {author} {\bibfnamefont {D.~H.}\ \bibnamefont {Richter}},
  \bibinfo {author} {\bibfnamefont {E.~S.~G.}\ \bibnamefont {Shaqfeh}}, \ and\
  \bibinfo {author} {\bibfnamefont {S.}~\bibnamefont {Chu}},\ }\href {\doibase
  10.1021/ma070636b} {\bibfield  {journal} {\bibinfo  {journal}
  {Macromolecules}\ }\textbf {\bibinfo {volume} {40}},\ \bibinfo {pages} {2461}
  (\bibinfo {year} {2007})}\BibitemShut {NoStop}%
\bibitem [{\citenamefont {Pan}\ \emph {et~al.}(2014)\citenamefont {Pan},
  \citenamefont {Nguyen}, \citenamefont {Sridhar}, \citenamefont {Sunthar},\
  and\ \citenamefont {Prakash}}]{Pan2014}%
  \BibitemOpen
  \bibfield  {author} {\bibinfo {author} {\bibfnamefont {S.}~\bibnamefont
  {Pan}}, \bibinfo {author} {\bibfnamefont {D.~A.}\ \bibnamefont {Nguyen}},
  \bibinfo {author} {\bibfnamefont {T.}~\bibnamefont {Sridhar}}, \bibinfo
  {author} {\bibfnamefont {P.}~\bibnamefont {Sunthar}}, \ and\ \bibinfo
  {author} {\bibfnamefont {J.~R.}\ \bibnamefont {Prakash}},\ }\href {\doibase
  10.1122/1.4861072} {\bibfield  {journal} {\bibinfo  {journal} {Journal of
  Rheology}\ }\textbf {\bibinfo {volume} {58}},\ \bibinfo {pages} {339}
  (\bibinfo {year} {2014})}\BibitemShut {NoStop}%
\bibitem [{\citenamefont {Graessley}(1980)}]{Graessley1980}%
  \BibitemOpen
  \bibfield  {author} {\bibinfo {author} {\bibfnamefont {W.~W.}\ \bibnamefont
  {Graessley}},\ }\href {\doibase 10.1016/0032-3861(80)90266-9} {\bibfield
  {journal} {\bibinfo  {journal} {Polymer}\ }\textbf {\bibinfo {volume} {21}},\
  \bibinfo {pages} {258} (\bibinfo {year} {1980})}\BibitemShut {NoStop}%
\bibitem [{\citenamefont {Tanyeri}\ and\ \citenamefont
  {Schroeder}(2013)}]{Tanyeri2013}%
  \BibitemOpen
  \bibfield  {author} {\bibinfo {author} {\bibfnamefont {M.}~\bibnamefont
  {Tanyeri}}\ and\ \bibinfo {author} {\bibfnamefont {C.~M.}\ \bibnamefont
  {Schroeder}},\ }\href@noop {} {\bibfield  {journal} {\bibinfo  {journal}
  {Nano Letters}\ }\textbf {\bibinfo {volume} {13}},\ \bibinfo {pages} {2357}
  (\bibinfo {year} {2013})}\BibitemShut {NoStop}%
\bibitem [{\citenamefont {Perkins}\ \emph {et~al.}(1997)\citenamefont
  {Perkins}, \citenamefont {Smith},\ and\ \citenamefont {Chu}}]{Perkins1997a}%
  \BibitemOpen
  \bibfield  {author} {\bibinfo {author} {\bibfnamefont {T.~T.}\ \bibnamefont
  {Perkins}}, \bibinfo {author} {\bibfnamefont {D.~E.}\ \bibnamefont {Smith}},
  \ and\ \bibinfo {author} {\bibfnamefont {S.}~\bibnamefont {Chu}},\ }\href
  {\doibase 10.1126/science.276.5321.2016} {\bibfield  {journal} {\bibinfo
  {journal} {Science}\ }\textbf {\bibinfo {volume} {276}},\ \bibinfo {pages}
  {2016} (\bibinfo {year} {1997})}\BibitemShut {NoStop}%
\bibitem [{\citenamefont {Zhou}\ and\ \citenamefont
  {Schroeder}(2016)}]{Zhou2016}%
  \BibitemOpen
  \bibfield  {author} {\bibinfo {author} {\bibfnamefont {Y.}~\bibnamefont
  {Zhou}}\ and\ \bibinfo {author} {\bibfnamefont {C.~M.}\ \bibnamefont
  {Schroeder}},\ }\href {\doibase 10.1103/PhysRevFluids.1.053301} {\bibfield
  {journal} {\bibinfo  {journal} {Physical Review Fluids}\ }\textbf {\bibinfo
  {volume} {1}},\ \bibinfo {pages} {053301} (\bibinfo {year}
  {2016})}\BibitemShut {NoStop}%
\bibitem [{\citenamefont {Robertson}\ and\ \citenamefont
  {Smith}(2007{\natexlab{b}})}]{Robertson2007a}%
  \BibitemOpen
  \bibfield  {author} {\bibinfo {author} {\bibfnamefont {R.~M.}\ \bibnamefont
  {Robertson}}\ and\ \bibinfo {author} {\bibfnamefont {D.~E.}\ \bibnamefont
  {Smith}},\ }\href {\doibase 10.1021/ma070051h} {\bibfield  {journal}
  {\bibinfo  {journal} {Macromolecules}\ }\textbf {\bibinfo {volume} {40}},\
  \bibinfo {pages} {3373} (\bibinfo {year} {2007}{\natexlab{b}})}\BibitemShut
  {NoStop}%
\bibitem [{\citenamefont {Gong}\ and\ \citenamefont {{Van Der
  Maarel}}(2014)}]{Gong2014}%
  \BibitemOpen
  \bibfield  {author} {\bibinfo {author} {\bibfnamefont {Z.}~\bibnamefont
  {Gong}}\ and\ \bibinfo {author} {\bibfnamefont {J.~R.~C.}\ \bibnamefont {{Van
  Der Maarel}}},\ }\href {\doibase 10.1021/ma501618a} {\bibfield  {journal}
  {\bibinfo  {journal} {Macromolecules}\ }\textbf {\bibinfo {volume} {47}},\
  \bibinfo {pages} {7230} (\bibinfo {year} {2014})}\BibitemShut {NoStop}%
\bibitem [{\citenamefont {Samsal}\ \emph {et~al.}(2017)\citenamefont {Samsal},
  \citenamefont {Hsiao}, \citenamefont {Schroeder},\ and\ \citenamefont
  {Prakash}}]{Samsal2017}%
  \BibitemOpen
  \bibfield  {author} {\bibinfo {author} {\bibfnamefont {C.}~\bibnamefont
  {Samsal}}, \bibinfo {author} {\bibfnamefont {K.-W.}\ \bibnamefont {Hsiao}},
  \bibinfo {author} {\bibfnamefont {C.~M.}\ \bibnamefont {Schroeder}}, \ and\
  \bibinfo {author} {\bibfnamefont {J.~R.}\ \bibnamefont {Prakash}},\ }\href
  {\doibase 10.1122/1.4972237} {\bibfield  {journal} {\bibinfo  {journal}
  {Journal of Rheology}\ }\textbf {\bibinfo {volume} {61}},\ \bibinfo {pages}
  {169} (\bibinfo {year} {2017})}\BibitemShut {NoStop}%
\bibitem [{\citenamefont {Robertson}\ \emph {et~al.}(2006)\citenamefont
  {Robertson}, \citenamefont {Laib},\ and\ \citenamefont
  {Smith}}]{Robertson2006a}%
  \BibitemOpen
  \bibfield  {author} {\bibinfo {author} {\bibfnamefont {R.~M.}\ \bibnamefont
  {Robertson}}, \bibinfo {author} {\bibfnamefont {S.}~\bibnamefont {Laib}}, \
  and\ \bibinfo {author} {\bibfnamefont {D.~E.}\ \bibnamefont {Smith}},\ }\href
  {\doibase 10.1073/pnas.0601903103} {\bibfield  {journal} {\bibinfo  {journal}
  {Proceedings of the National Academy of Sciences of the United States of
  America}\ }\textbf {\bibinfo {volume} {103}},\ \bibinfo {pages} {7310}
  (\bibinfo {year} {2006})}\BibitemShut {NoStop}%
\bibitem [{\citenamefont {Bhattacharjee}\ \emph {et~al.}(2003)\citenamefont
  {Bhattacharjee}, \citenamefont {Nguyen}, \citenamefont {McKinley},\ and\
  \citenamefont {Sridhar}}]{Bhattacharjee2003}%
  \BibitemOpen
  \bibfield  {author} {\bibinfo {author} {\bibfnamefont {P.~K.}\ \bibnamefont
  {Bhattacharjee}}, \bibinfo {author} {\bibfnamefont {D.~A.}\ \bibnamefont
  {Nguyen}}, \bibinfo {author} {\bibfnamefont {G.~H.}\ \bibnamefont
  {McKinley}}, \ and\ \bibinfo {author} {\bibfnamefont {T.}~\bibnamefont
  {Sridhar}},\ }\href {\doibase 10.1122/1.1530625} {\bibfield  {journal}
  {\bibinfo  {journal} {Journal of Rheology}\ }\textbf {\bibinfo {volume}
  {47}},\ \bibinfo {pages} {269} (\bibinfo {year} {2003})}\BibitemShut
  {NoStop}%
\end{thebibliography}%
\end{document}